%% file: IEEE-conf-template.tex
\documentclass[conference]{IEEEtran}
\IEEEoverridecommandlockouts
\usepackage{multirow}
\usepackage{pifont}
\usepackage{caption}
\usepackage{subcaption}
\usepackage[htt]{hyphenat}
\usepackage[hyphenbreaks]{breakurl}
\usepackage[hyphens]{url}
\usepackage{comment}
\usepackage{cite}
\usepackage{amsmath,amssymb,amsfonts}
\usepackage{listings}
\usepackage{color}
\definecolor{backcolour}{rgb}{0.95,0.95,0.92}
\definecolor{codepurple}{rgb}{0.58,0,0.82}
\definecolor{numbercolor}{rgb}{0.205, 0.142, 0.73}

\usepackage{algorithm}
\usepackage{algpseudocode}

\usepackage{graphicx}
\usepackage{textcomp}
\usepackage{xcolor}
\def\BibTeX{{\rm B\kern-.05em{\sc i\kern-.025em b}\kern-.08em
    T\kern-.1667em\lower.7ex\hbox{E}\kern-.125emX}}
\begin{document}

\title{Few-Shot Learning-Based Cyber Incident Detection with Augmented Context Intelligence}
%Towards Imbalanced Events in Cloud Threats

%\author{Anonymous Submission}
\include{author}

\maketitle

\begin{abstract}

In recent years, the adoption of cloud services has been expanding at an unprecedented rate. As more and more organizations migrate or deploy their businesses to the cloud, a multitude of related cybersecurity incidents such as data breaches are on the rise. Many inherent attributes of cloud environments, for example, data sharing, remote access, dynamicity and scalability, pose significant challenges for the protection of cloud security. Even worse, cyber threats are becoming increasingly sophisticated and covert. Attack methods, such as Advanced Persistent Threats (APTs), are continually developed to bypass traditional security measures. Among the emerging technologies for robust threat detection, system provenance analysis is being considered as a promising mechanism, thus attracting widespread attention in the field of incident response. 
This paper proposes a new few-shot learning-based attack detection with improved data context intelligence. We collect operating system behavior data of cloud systems during realistic attacks and leverage an innovative semiotics extraction method to describe system events. Inspired by the advances in semantic analysis, which is a fruitful area focused on understanding natural languages in computational linguistics, we further convert the anomaly detection problem into a similarity comparison problem. Comprehensive experiments show that the proposed approach is able to generalize over unseen attacks and make accurate predictions, even if the incident detection models are trained with very limited samples.

\end{abstract}

\begin{IEEEkeywords}
Incident Detection, Anomaly Detection, Cyber Threat, Cloud Security, Few-Shot Learning.
\end{IEEEkeywords}
%Incident Response

\input{intro}

\input{background}

\input{define}

\input{design}

\input{eval}

\input{relate}

\input{conclude}

\input{ack}

\bibliographystyle{IEEEtran}
\bibliography{refs}

\end{document}

%% file: author.tex
\author{\IEEEauthorblockN{
Fei Zuo$^*$, Junghwan Rhee$^*$, Yung Ryn Choe$^\dag$, Chenglong Fu$^\ddag$, Xianshan Qu$^*$}
\IEEEauthorblockA{
$^*$University of Central Oklahoma, $^\dag$Sandia National Laboratories, $^\ddag$University of North Carolina at Charlotte}
\{fzuo, jrhee2, xqu1\}@uco.edu, yrchoe@sandia.gov, chenglong.fu@uncc.edu
}

\begin{comment}

\author{\IEEEauthorblockN{1\textsuperscript{st} Given Name Surname}
\IEEEauthorblockA{\textit{dept. name of organization (of Aff.)} \\
\textit{name of organization (of Aff.)}\\
City, Country \\
email address or ORCID}
\and
\IEEEauthorblockN{2\textsuperscript{nd} Given Name Surname}
\IEEEauthorblockA{\textit{dept. name of organization (of Aff.)} \\
\textit{name of organization (of Aff.)}\\
City, Country \\
email address or ORCID}
\and
\IEEEauthorblockN{3\textsuperscript{rd} Given Name Surname}
\IEEEauthorblockA{\textit{dept. name of organization (of Aff.)} \\
\textit{name of organization (of Aff.)}\\
City, Country \\
email address or ORCID}
\and
\IEEEauthorblockN{4\textsuperscript{th} Given Name Surname}
\IEEEauthorblockA{\textit{dept. name of organization (of Aff.)} \\
\textit{name of organization (of Aff.)}\\
City, Country \\
email address or ORCID}
\and
\IEEEauthorblockN{5\textsuperscript{th} Given Name Surname}
\IEEEauthorblockA{\textit{dept. name of organization (of Aff.)} \\
\textit{name of organization (of Aff.)}\\
City, Country \\
email address or ORCID}
\and
\IEEEauthorblockN{6\textsuperscript{th} Given Name Surname}
\IEEEauthorblockA{\textit{dept. name of organization (of Aff.)} \\
\textit{name of organization (of Aff.)}\\
City, Country \\
email address or ORCID}
}
\end{comment}

%% file: intro.tex
\section{Introduction}

Cybersecurity incidents are emerging incessantly nationwide. Commercial organizations, government bodies, and even educational institutions can all be potential targets for cyberattacks. For example, an investigation~\cite{ibm2023} conducted on 550 organizations revealed that 83\% of them had more than one data breach, and the cost of a data breach averaged USD 4.35 million. Furthermore, over the recent years, a noteworthy trend has been the sharp increase in the number of attacks on cloud environments per organization, ``\textit{which shot up by 48\% in 2022 compared with 2021}''~\cite{check2023}. Even worse, cloud environments are demonstrably vulnerable to Advanced Persistent Threats (APTs), which can be covert over a prolonged period of time but difficult to defend. Therefore, in this paper, we focus on threat detection towards cloud incidents.

In the arsenal of threat detection, system provenance analysis is believed to possess great potential in detecting cyber threats because system-level data can not only represent complex dependencies within a system, which is crucial for understanding potential threats, but also correlate with attack scenarios, providing a valuable historical context that helps in predicting and preventing future attacks. The two interrelated system entities, along with the operation between them, collectively constitute an event, which is a basic unit in provenance data for tracking and recording system-level behaviors. We thus extract events from a provenance graph and regard them as informative features for describing the characteristics of a cyber incident.

More specifically, in an event, the entity that issues an operation is modeled as a \textit{subject}, while the other entity that passively undergoes an operation is modeled and referred to as an \textit{object}. Inspired by this, we analogize the semantic description of an event to a \textit{sentence} in linguistics. To this end, we particularly develop a semiotics extraction method to capture the event's semantics with enriched expressivity. Then, we further adopt the embedding method in Natural Language Processing (NLP) to generate a numeric representation for every event's description, which will lay a solid foundation for the subsequent threat detection. 

Furthermore, we notice that with the tremendous success of machine learning in the recent decade, leveraging related advancements to assist and automate system provenance analysis has become more of a need than an option. Industry statistics show that in practical incident response applications, fully deployed AI-driven systems ``\textit{were able to identify and contain a breach 28 days faster than those that didn’t, saving USD 3.05 million in costs}'' for the organizations~\cite{ibm2023}. 

It is generally acknowledged that training high-quality machine learning models typically requires massive amounts of data. However, well-labeled large datasets are usually a precious or even scarce resource in the field of cybersecurity. 
%~\cite{Shrestha2023provsec,zuo2023secdb}
One widely seen challenge in collecting cyber incident data is that the actually expected distribution of samples is not balanced because system events in practical scenarios are mostly benign. Even worse, this imbalance is quite likely to result in a high false positive rate, which would be less useful for many security-sensitive tasks. Not surprisingly, researchers consider existing threat detection methods ``\textit{may not be robust enough to distinguish malicious behavior from benign ones accurately}''~\cite{li2021threat}. Thus, what we are looking for is a more effective threat detection technique towards realistic incidents in cloud environments, which can cope with not only various known attacks but also previously unseen ones.

%C02:Apache2 path traversal exploitation, 
\begin{figure*}[htbp]
  \centering
  \includegraphics[width=0.72\textwidth]{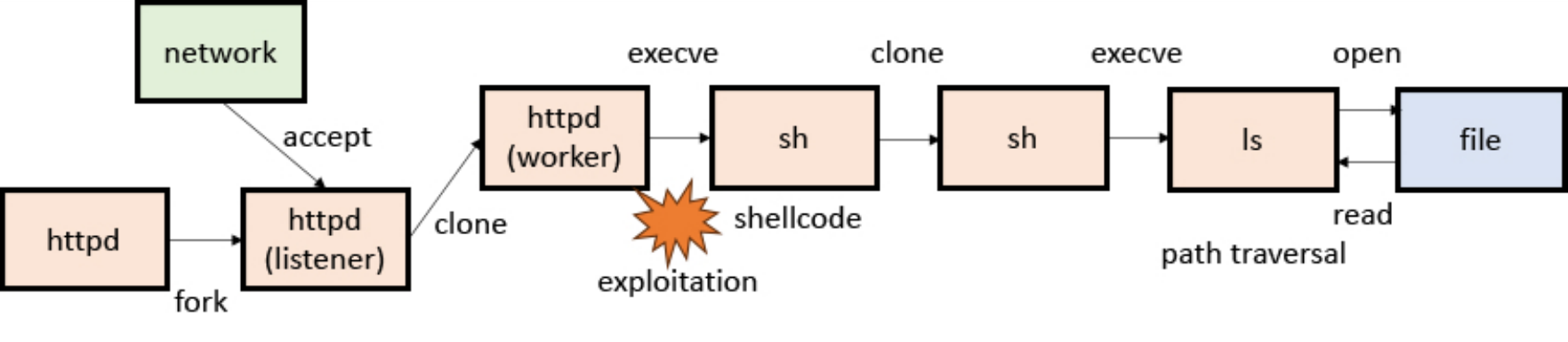}
  \caption{A provenance graph example (CVE-2021-41773 case).}
  \label{fig:prov_graph}
\end{figure*}

In the field of machine learning, Few-Shot Learning (FSL) aims at enabling models to generalize over unseen data and make accurate predictions, even if they are trained with very limited samples. In light of the great progress in FSL, we convert anomaly detection within the context of incident response into a semantic comparison problem. Our intuition is that normal and attack behaviors can be mapped to two different clusters in a certain feature space. The FSL-based neural network model is thus used as a kernel function, which maps the data points from the original space to a feature space where the intra-cluster distance is relatively larger than the inter-cluster distance. Based on a well-trained model, we are able to predict whether a previously unseen behavior is an attack or not, by comparing it with a known case. 

Following the aforementioned pipeline, we develop a cyber threat detection system for incident response. 
First, we adopt an innovative semiotic extraction method, which can accurately capture the key semantics of system call-based program behaviors. To generalize cyber threat intelligence over unseen attacks, we present a new few-shot learning technique applied to provenance data, which converts an anomaly detection problem into a similarity comparison problem. 
We have explored a systematic strategy to address security threats in an increasingly challenging cloud environment. To this end, realistic cyber-attacks on cloud applications are investigated in this work.
At last, wee empirically demonstrated that few-shot learning is applicable to  operating system behavior datasets. The comprehensive experiments show that our proposed approach is capable of generalizing over previously unseen attacks and making accurate classifications, even if there exist very limited training samples in each attack scenario.

%% file: background.tex
\section{Background}\label{sec:bck}
\begin{comment}
In this section, we briefly introduce the necessary background on 
%system-level 
provenance data, APT attacks in cloud environments, and few-shot learning.
%that are involved in this paper.
\end{comment}

%\subsection{Provenance Data}

In incident response research and practice, provenance data reflects activities at the level of system entities,
%or syscalls, 
which offers great potential for tracking the dependencies across system events and further exposing attack sequences. 
System-level provenance can usually be recorded in the form of graphs, where vertices and edges represent entities and the operations amongst entities respectively. 

Figure~\ref{fig:prov_graph} shows a provenance graph example that exposes an attack scenario in an Apache HTTP server. Upon a connection from the network with an {\small\texttt{accept}} system call, a worker process of the web server ({\small\texttt{httpd}}) is invoked. Then due to the exploit code from an attack, this server allows the shellcode which invokes the {\small\texttt{ls}} process that lists files as a demonstration of arbitrary command execution.

The system entities refer to the components within a system responsible for producing, modifying, or processing information and resources. Examples of entities include processes, files, and network objects, which are indicated by orange, blue, and green nodes in Figure~\ref{fig:prov_graph}, respectively. 
For various operating systems, similar types of entities can be found. Therefore, 
we can apply a similar approach to different operating systems.
%it is not hard to extend the provenance graph with more types of entities targeting different operating systems. 

The operations between two system entities are described using a system call, which is a lower-level service interface
invoked by software to use the privileged services or resources provided by the kernel. For instance, process operations can include {\small\texttt{execve}}, {\small\texttt{fork}} and {\small\texttt{clone}} system calls. Besides, {\small\texttt{open}}, {\small\texttt{close}}, \texttt{read}, and \texttt{write} system calls are examples of file operations. Lastly, operations such as {\small\texttt{connect}} and {\small\texttt{accept}} system calls are invoked by network activities. 

When taking both operations and the entities that issue such operations into consideration, all these elements come together to form an event. Moreover, multiple events can manifest causality dependency if there exist direct or indirect data and control flows across them. Provenance data is often represented as a graph because graphs can intuitively depict dependencies between events or their chronological order. However, considering
that hackers launch multi-stage attacks, extracting attack-relevant events from a vast number of system events over a long time span
and linking them is non-trivial. As a result, the generated provenance graph often contains significant noise, posing challenges for
threat detection. Therefore, it should be noted that \textbf{the proposed approach focuses on system events in provenance data}.

%% file: define.tex
\section{Problem Definition and Basic Assumptions}\label{sec:prb}

%In this section, we formulate the problem statement and present the assumptions of attack scenarios. 

\subsection{Problem Statement and Objectives}

In the operating system of a cloud environment, the target that we are interested in monitoring is processes. A process $p$ is an executing instance of a program $A$. The behavior of a process in the system is recorded in the form of provenance data consisting of a large number of events. Given a new process instance $p$ of $A$, our aim is to infer whether it implicitly involves a malicious attack by analyzing its provenance data.
%we aim to detect whether its provenance graph represents malicious attacks. 
In particular, our research objective is to seek an effective threat detection technique for realistic incidents in cloud, which can handle previously unseen attacks. 
Intuitively, capturing and modeling the behaviors of each program by a provenance graph seems feasible, but challenges still lie ahead of us. In practice, we need to address the following two challenges.

\begin{figure*}[ht]
  \centering
  \includegraphics[width=0.72\textwidth]{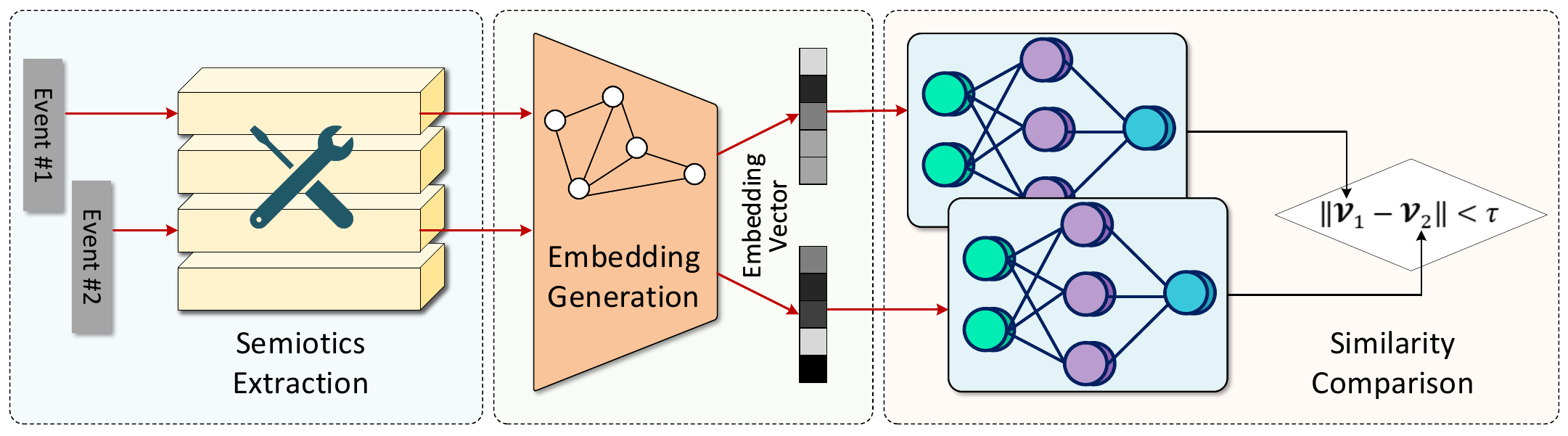}
  \caption{System overview.}
  \label{fig:overview}
\end{figure*}

\textbf{Challenge 1: Representativeness and Generalizability.} In modern computer systems, the programs that might be used are diverse. Exhaustively enumerating every possible program and empirically constructing provenance graphs to indicate the presence or absence of malicious attacks requires domain expertise and substantial human effort. 
In addition, cloud environments are complex and dynamic. For the same program, provenance graphs from a similar workload may look identical, but if the input workloads differ, the resulting provenance graphs may also differ. Therefore, it is uncertain whether stereotype provenance graphs established based on experience or limited simulations can cover all possible scenarios.
Finally, malicious attacks can also be mixed and hidden in normal operations. Therefore, attackers can generate a complex provenance graph by adding a large number of normal events, attempting to obscure the malicious events within. As a result, treating the graph as a whole for anomaly detection becomes undesirable.

To address this challenge, we decompose a provenance graph into events, which is a basic unit for recording system-level behaviors. The benefits of this approach are evident. Different APT attacks in a cloud environment share commonalities at the level of system calls. 
%They can be typically encapsulated by the four stages 
%depicted in Figure~\ref{fig:apt_attack}. 
%This point has been consistently summarized and confirmed by previous research~\cite{swathy2019taxonomy, choi2015ontology}. 
Therefore, malicious events captured in one attack scenario can be generalized to other unseen attack scenarios. 
Again, from the attack example of Apache HTTP server (Figure~\ref{fig:prov_graph}), 
%and Log4j (Figure~\ref{fig:apt_attack}), 
we observed that remote shell execution is issued by {\small\texttt{httpd}}, 
%and \texttt{java} respectively, 
indicating exploitation. However, these events should not occur when normal workloads are input. 
%For instance, \texttt{java} does not run any other software as a child when involving a java-based logging utility. 
On top of that, the object of examination is no longer the entire graph, but individual events. Hence, methods that use normal operations to obscure malicious attacks at the graph level are no longer effective. %Our subsequent visualization in Section~\ref{sec:eval}.4 also demonstrates that, 
After using appropriate embedding methods, benign events form clusters in the high-dimensional feature space, which are distinct from the clusters formed by adversarial events.

\textbf{Challenge 2: Feature Preparation for Anomaly Detection.} The performance of traditional machine learning methods heavily relies on feature engineering, where domain knowledge plays a significant role in creating and selecting features from raw data. Moreover, current powerful system provenance data recorders, e.g., {\small\texttt{Sysdig}} and {\small\texttt{Auditd}}, are able to provide abundant fields to describe an event, for instance, but not limited to timestamps, system call names, process IDs (PIDs), file names, IP addresses, port numbers, and user IDs (UIDs), etc. Therefore, which features to select and how to evaluate the impact of each feature on the performance of anomaly detection in different attack scenarios depend on expertise and manual effort. Not only that, many fields in provenance data are descriptive text and therefore unstructured, such as file paths or process names. These unstructured data can be further divided into two categories. The first category consists of meaningful entities that correspond to specific tasks in the system.
%, such as \texttt{java} and \texttt{sh} in Figure~\ref{fig:apt_attack}. 
The other category consists of non-representative entities that are temporarily generated during intermediate steps, such as temporary files. Obviously, we need to design different response strategies for these two categories of entities. To sum up, how to prepare input features to represent system events for anomaly detection models is non-trivial.

To address this challenge, we view a system event as a sentence in natural language. The nodes and edges in a graph correspond to the subject, predicate, object, or complement of a sentence, which describes a system-level behavior. Furthermore, we can project a sentence to a numerical vector in the feature space using those advanced neural networks-based embedding methods. A key advantage of neural networks is their ability to learn directly from the raw data with minimal need for feature engineering. Relying on the embedding method for accurate semantic capture, in the feature space, sentences with similar semantics have closer vector representations, while different sentences are farther apart. For example, ``{\small\texttt{java execve sh}}'' is supposed to possess a numerical vector representation approximating that of ``{\small\texttt{httpd execve sh}}''. As a result, we are able to develop distance-based classifiers as our anomaly detection models. 
%Furthermore, the number of adversarial and benign events is often imbalanced in reality. However, embedding techniques are self-supervised, which means that even if we do not have enough adversarial events, we can learn vector encoding methods using purely benign data. 
%We detail our proposed method in Section \ref{sec:method}.3.

\subsection{Threat Model and Assumptions}

Our threat model has several assumptions on adversary activities of our interest. First, we assume the security attack involves the change of the operating system calls as behavior to use system resources or services. While it is possible to make trivial attacks without involving system calls, they would have very limited impact because most serious impacts such as privilege changes, files, network activities seen in privilege escalation and data leak require system calls. The attacks without any system calls are hence out of focus. 

\begin{table*}[!th]
\caption{Types of system calls used for semiotics.}\label{tab:syscalls}
\centering
\begin{center}
\begin{tabular}{ l |l  }
\hline
 Event type & System call names  \\ 
 \hline
 Process events  & \texttt{fork, vfork, clone, exec, kill}, and variants \\  
 File events  & \texttt{open, close, read, write, unlink, dup, rename, chmod}, and variants \\
 Network events  & \texttt{listen, connect}, and variants \\   
 \hline
\end{tabular}
\end{center}
\end{table*}

Second, adversaries may use software exploits to cause compromise of software and subsequent damages. However, we assume the integrity of the security event recorder so that the system calls of the adversary events can be properly recorded.
Similar to event detection and response (EDR) solutions, the recorded events are remotely transferred and the integrity of the OS kernel and the data recorder are assumed to be protected.

%Third, as a machine learning-based approach we assume the security defenders have behavior data similar to potential adversary attacks. This is a reasonable assumption applied to any machine learning-based security work because the attack behavior mechanism is well documented in the community as evidenced by the MITRE's ATT\&CK framework~\cite{strom2018mitre}. Particularly, we use an open security behavior dataset [Hidden due to blind review] for training and testing for our prototype.
%The reason why we chose this dataset for our prototype is that it has high-quality labels with clear ground truths. In addition, since we are the authors who produced benign and adversary cases ourselves in the cloud environment with real exploits, we are very knowledgeable about the details of the data.
%Note our proposed approach is general to work with any properly labeled system call behavior datasets and a larger scale evaluation with high-quality datasets is our future work. 
%The strength of the mechanism can be enlarged by using a larger set of security datasets available in the community. 
%Unknown or zero-day attacks whose behavior cannot be categorized with prior attack data are not our focus.

%% file: design.tex
\section{Methodology}\label{sec:method}

In this section, we propose a new cyber
threat intelligence framework for the analysis and detection of cloud incidents. Figure~\ref{fig:overview} shows the overview of our proposed approach, which consists of three major components, i.e., semiotics extraction, embedding generation, and similarity comparison. 
%We first introduce our key insights. Based on them, 
%we explain the intelligence-augmented method proposed by us for robust and universally applicable threat detection in cloud environments.

\subsection{Key Insights}

Our proposed methods are motivated by several key insights on the provenance data collected from realistic cloud incidents.

\textbf{Program behavior can be abstracted into text}. Programs utilize common OS interfaces that have a uniform format. Hence, we can standardize the description of program behavior into generic sentences.

\textbf{\textit{Insight} 1:} We embed system event details as natural language sentences that are composed of subject entities (i.e., programs), the type of operations acted by the subject entities as a predicate (e.g., system calls), and object entities that are the target of the operations by the subject entities. The object component includes \ding{182} processes, \ding{183} files, and \ding{184} network entities (e.g., IP addresses).

\textbf{Certain process events lack sufficient details}. Certain system events may have limited context of behavior because of the way software handles code. One example is interpreters such as {\small\texttt{bash}}, {\small\texttt{python}}, and {\small\texttt{java}}. These programs rely on  common software binaries even though the actual software code differs. Such differences in program code are not determined from the main executable names (i.e., interpreters' names) because they are provided as parameters to them. Our insight was that assisting the model in such lacking contexts can drastically improve the performance of machine learning based solutions.

\textbf{\textit{Insight} 2:} In linguistics, an object complement is a grammatical construction that provides additional information about a direct object in a sentence. 
Drawing an analogy to this, we improved the system events description by supplementing extra information such as process executable names.

\textbf{Non-representative files are widespread.}
Operating systems use a lot of ``temporarily generated'' information, which is used in intermediate steps. Such information is generally not important, as it is eventually deleted after a certain period or upon reboot, and it can introduce unnecessary complexity into the machine learning process. These files can be identified due to specific locations or naming schemes.
%We create algorithms specific to operating system calls that can further strengthen the machine learning models by identifying and handling these files. 

\textbf{\textit{Insight} 3:} We leveraged multiple patterns to recognize and normalize non-representative files. In addition, we use an algorithm to detect hash-like names based on a high diversity of file names.

\begin{comment}

\textbf{Insight 2: Similarity in symptoms across attacks based on OS level abstractions.}
Few-shot learning algorithms are applied to instances whose samples are lacking. This would be challenging for machine learning if the models cannot learn certain types of behavior that would be applicable to new cases. Different attacks may utilize different types of software vulnerabilities (e.g., SQL injection vs. Buffer overflow) and different software logic (e.g., Apache webserver vs. Nginx webserver). However, the program behavior that we use is described using an operating system abstraction that can extract and recognize similarities across programs and attacks. For instance, behaviors of path traversal attacks targeting different webserver software are similar in making system calls as they need to use the same directory open syscalls. As another example, adversary behaviors such as backdoors show similar behavior if they share the OS input and output behavior despite different codes for implementation.

\textbf{Method from Insight 2: Few-shot learning.}
Numerous attack mechanisms that have been developed and it is very challenging to create a customized detector for every individual attack mechanism. Instead, we would like to utilize the few-shot learning technique so that we can improve the applicability of our incident detection mechanism to a wider variety of attacks.

\end{comment}

\subsection{Semiotics Extraction}

%Herein, we provide more technical details, so that other researchers can follow the same procedures to reproduce our cyber threat intelligence.

We use operating system calls
%~\cite{silberschatz2018operating} 
for our behavior modeling. While our experiment is mainly focused on Linux, our method is general and the same mechanism can be applied to other operating systems which also have system calls as seen in related work \cite{priotracker, nodoze, 9152771, 10.1145/3243734.3243763}.

\subsubsection{Program behavior abstraction}
% \textbf{Collected system call types.} 
Table \ref{tab:syscalls} shows a list of system calls collected, extracted, and used in our work. Process events include process creation system calls such as {\small\texttt{fork}}, {\small\texttt{vfork}}, and {\small\texttt{clone}} and the {\small\texttt{exec}} system call that replaces the program image inside a process is also used. {\small\texttt{kill}} system calls are used to terminate processes. File events refer to file behavior such as file creation, file deletion, file open, file close, file read, file write, rename, duplication etc. For instance, {\small\texttt{open}}, {\small\texttt{close}}, {\small\texttt{read}}, {\small\texttt{write}} are common examples.
In Unix-like operating systems, almost everything is accessible as a file. Therefore, most system behavior can be tracked by utilizing file activities. Over the years, operating systems have added support for variant system calls. One example is the {\small\texttt{*at}} system calls (e.g., {\small\texttt{openat}}) which enable file operations relative to the referred directory descriptor in a parameter. Another example is {\small\texttt{p*}} system calls (e.g., {\small\texttt{popen}}) which are used to operate a process with a pipe. We only list the major system calls here. Their variants are available depending on the versions of the Linux kernels. Network events for both the client side (e.g., {\small\texttt{connect}}) and the server side (e.g., {\small\texttt{listen}}) are considered.

\textbf{Modeling system events as sentences.}
%Linux system calls are commonly executed in the following format. System call names are invoked either by the user code directly (as seen in \texttt{fork} or \texttt{kill}) or via the library code which uses system calls internally. For instance, memory allocation and release can be performed using \texttt{malloc} and \texttt{free}, and they internally use system calls to allocate and adjust memory allocated to a process using the \texttt{brk} and/or \texttt{sbrk} memory management system calls. 
A system call is always executed on behalf of a particular process, which serves as the identity of that particular system call action. Hence, we model a system event in the following triplet form, that is composed of a {\small\texttt{Subject}} (a process initiating a system call), {\small\texttt{Predicate}} (a system call type), and {\small\texttt{Object}} (a target upon which the system call is applied). For instance, in the example below, we have two system event descriptions where the {\small\texttt{ps}} program operates on the \texttt{stat} file.
\begin{small}
\begin{verbatim}
(Subject)  (Predicate) (Object)
 ps         open        stat
 ps         close       stat
\end{verbatim}
\end{small}
%This approach is similar to what related approaches have used \cite{Wang2020YouAW,sigl}. Now we would like to introduce what kind of data processing we have used to assist Few-Shot Learning.

\textbf{Data enhancement for certain program types.} 
%During our experiments, we find that certain types of programs show less performance than others. 
In Unix-like operating systems, a program identity is determined by an executable loaded by the {\small\texttt{exec}} system call.
%of operating systems. 
While this method works for most programs, there are exceptions where program executables are not representative. 

\begin{enumerate}
    \item \textbf{Interpreter-based scripting shells:} Shells with scripting capabilities use the script interpreter as the main executable. For example, Linux shells (e.g., {\small\texttt{bash}}, {\small\texttt{csh}}, {\small\texttt{zsh}}) use the main script file as the first parameter allowing it to be used together with the executable file to represent the program.
    \item \textbf{Software platforms using interpreters:} Multiple software languages use interpretation for code execution. Well known examples include {\small\texttt{Python}}, {\small\texttt{Perl}}, {\small\texttt{Ruby}}, and {\small\texttt{JavaScript}}. Programs falling into this category will be recognized by its interpreter executable name and the script file name.
    \item \textbf{Single software-based\footnote{We used the term, ``single software-based'' virtual machine to differentiate it from the virtual machines that execute an OS such as \texttt{virtualbox}.} virtual machines:} Virtual machines execute an executable after obtaining the code file as one of the command-line parameters. For instance, {\small\texttt{Java}}, {\small\texttt{Scala}}, and {\small\texttt{Kotlin}} use the Java Virtual Machine (JVM) to run their executable files.
\end{enumerate}
%These cases commonly describe the identity of the program as one of the parameters that refer to the program file. 
These cases commonly describe the program's identity as one of the parameters referring to the program file. Consequently, we handled such cases by appending the program file parameter to the behavior description sentence, which is analogical to an object complement in a natural language sentence.

\textbf{New modeling of system events as sentences.} Based on the initial modeling introduced earlier, we further improve it with the following new ideas in this work. The new augmented sentence format thus obtained is 
\begin{small}
\begin{verbatim}
(Subject) (Predicate) (Subject complement)
(Object) (Object complement)
\end{verbatim}
\end{small}

%\noindent{\small\texttt{(Subject)(Predicate)(Subject complement)}}
%\noindent{\small\texttt{(Object)(Object complement)}}
\noindent This new sentence format is grounded in multiple aforementioned insights. The ``{\small\texttt{Subject complement}}'' and ``{\small\texttt{Object complement}}'' are two newly added optional components.
\begin{itemize}

    \item \textbf{Subject complement:} The data enhancement for shells, interpreters, and single-software based virtual machines are supported with supplemental parameters of program execution. Such information is listed as one of the subject complement after applying multiple normalization techniques. Plus, we simplify the program arguments by removing flags and normalizing hash-like values.
    
    \item \textbf{Object complement:} We applied normalization to the object of each system event as shown in Algorithm \ref{alg:normalization} (See Section~\ref{sec:normalize}). It uses multiple normalization techniques including hash detection to reduce the diversity of tokens that are not likely to be reused.
    
\end{itemize}

More concretely, we present several examples demonstrating how this augmented semantic information improves system event representations.

\begin{lstlisting}[
  label={list_eg01},
  caption={Example 1 of augmented system event expression.},
  captionpos=b,
  frame=shadowbox, 
  basicstyle          =   \ttfamily\footnotesize,
  rulesepcolor=\color{red!20!green!20!blue!20},
  keywordstyle=\bfseries\color{numbercolor},
  showstringspaces=false,
  breakatwhitespace=false,         
  breaklines=true,       
  keepspaces=true,
  xleftmargin=4pt,
  xrightmargin=6pt,
  morekeywords={BEFORE, AFTER}
]
# (Subject): java
# (Predicate): clone
# (Subject complement): executable parameter 
   extension = java 8983 /opt/solr/server/logs 
   start.jar
# (Object): <NA>

BEFORE: java clone <NA>
AFTER:  java clone java 8983 /opt/solr/server/logs
        start.jar <NA> 
\end{lstlisting}

Listing~\ref{list_eg01} shows the first example, where the augmented subject complement adds new information to the process context. The initial sentence ({\small\texttt{BEFORE}}) only had the program name ({\small\texttt{java}}), the system call name {\small\texttt{clone}}, and invalid target, {\small\texttt{<NA>}}, indicating that object information is not used for this type of system call ({\small\texttt{clone}}). The new format ({\small\texttt{AFTER}}) shows that, ffter filtering out the option fields starting with `{\small\texttt{-}}', several command-line parameters are listed, and the main Java JAR file, {\small\texttt{start.jar}}, appears.

%was initiated by the {\small\texttt{systemd}} process. 
%This program is executed by the \texttt{root} whose shell is \texttt{/bin/bash}. 

\begin{lstlisting}[
  label={list_eg02},
  caption={Example 2 of augmented system event expression.},
  captionpos=b,
  frame=shadowbox, 
  basicstyle          =   \ttfamily\footnotesize,
  rulesepcolor=\color{red!20!green!20!blue!20},
  keywordstyle=\bfseries\color{numbercolor},
  showstringspaces=false,
  breakatwhitespace=false,         
  breaklines=true,       
  keepspaces=true,
  xleftmargin=4pt,
  xrightmargin=6pt,
  morekeywords={BEFORE, AFTER}
]
# (Subject): dockerd
# (Predicate): openat
# (Object): .tmp-config.v2.json07205514
# (Object complement): 
#   .tmp-config.v2.json072055514 -> <TMP>

BEFORE: dockerd openat .tmp-config.v2.json07205514
AFTER:  dockerd openat <TMP>
\end{lstlisting}

As shown in Listing 2, the second example illustrates the normalization process. The original form ({\small\texttt{BEFORE}}) included a temporary file name, {\small\texttt{.tmp-config.v2.json07205514}}, which is not likely to be reused and thus is not helpful for subsequent machine learning task. The new format ({\small\texttt{AFTER}}) shows that the Docker daemon ({\small\texttt{dockerd}}) opens a temporary file that is normalized to {\small\texttt{<TMP>}}.

\subsubsection{Data normalization of non-representative data}\label{sec:normalize}
Modeling raw data requires careful arrangement of the data because it is well-known that non-representative (e.g., volatile noise that may not be used twice) can cause undesired side effects to machine learning performance.
For example, the exact names are not likely to appear in the test run. The resulting out-of-vocabulary issue often causes language models to misunderstand or misinterpret the meaning of the input. Therefore, these identifiers are better to be normalized for our machine-learning tasks. To this end, we propose to use several mechanisms based on reasonable assumptions.
We recognize such files and replace them using a constant label such as {\small\texttt{<TMP>}}, {\small\texttt{<PIPE>}}, or {\small\texttt{<HASH>}} depending on categories. Algorithm 1 specifically demonstrates our processing strategy.
%where {\small\texttt{IsHighVarietyCharacterName}} is a function to recognize names with a high variety of characters.

\input{pics/alg1}

\textbf{Normalization of non-representative files based on operating system knowledge:}
One assumption is that we can utilize well-known knowledge about operating systems. For instance, most operating systems have well-known common directory locations used by kernels (e.g., {\small\texttt{/bin}}, {\small\texttt{/tmp}}) as described in operating manuals. 
Certain directories in UNIX variants
such as {\small\texttt{/run/}}, {\small\texttt{/dev/}} and {\small\texttt{/proc/}} are all utilized to contain operating system internal states such as devices, and a list of program processes.
Such states have high volatility or randomness that may cause out-of-vocabulary issues in machine learning approaches.
Thus, the files in such directories are normalized by replacing them with {\small\texttt{<TMP>}}. Any file used for a pipe (in Unix everything is a file) is generalized as {\small\texttt{<PIPE>}} as well. 
For brand-new operating systems where pre-knowledge is incomplete, this knowledge can be easily figured out by a system administrator with a one-time effort.

\textbf{Normalization of temporary files:} Temporary files are generated by programs or operating systems for short-term use. They are another source of non-representative data, since such file names may not be persistent. Our assumption to handle this data is that such knowledge is also well-known or the knowledge can be easily obtained with one-time effort. For instance, {\small\texttt{/tmp}} in UNIX, {\small\texttt{C:\symbol{92}Users\symbol{92}Username\symbol{92}AppData\symbol{92}}}
{\small\texttt{Local\symbol{92}Temp}}, or {\small\texttt{C:\symbol{92}Windows\symbol{92}Temp}} in Windows are non-regular (i.e., specially designated) directories for temporary files. Algorithm \ref{alg:normalization} illustrates our practice in handling non-representative file names of targets whose names are randomly determined. 
In addition, the file extensions for temporary files (e.g., {\small\texttt{.tmp}}) are also used.

\textbf{Normalization of the identifiers with a high variety of characters:}
Another source of randomness comes from the identifiers that alternate numbers ({\small\texttt{0-9}}), alphabets ({\small\texttt{a-zA-Z}}), and special characters (e.g., {\small\texttt{@,-,\_,\$,\%}}). We observed that that certain programs generate hash-like filenames (e.g., {\small\texttt{75619cbc-879c-4076-8539-181392588ced}}) because they use certain algorithms such as UUID, MD5, SHA that alternate alphabets and numbers for filenames.  
%It can be seen that these hash-like filenames appear to be strings with a high variety of alternations across letters, digits, and special characters, which is also the fundamental characteristic of these filenames.
To mitigate the negative impact caused by non-representative data, any file or directory name recognized as a hash-like literal is normalized using {\small\texttt{<HASH>}}.
%(Assumption): Our assumption to address this challenge is that the names with a high variety of alternations across letters, digits, and special characters can be captured with a heuristic mechanism.
%with several thresholds on the variety of alternations.
%To mitigate the negative impact caused by non-representative data, we designed a heuristic hash detection algorithm {\small\texttt{IsHighVarietyCharacterName}}.
%Any file or directory name recognized with our algorithm is normalized as {\small\texttt{<HASH>}}.
%
%Algorithm  shows how we determine hash files statistically.
%This is to avoid the out-of-vocabulary problem, directory and the extension patterns are changed to include more information.
At a high level, our determination method uses generally two different strategies.
First, the files with well-known extensions such as {\small\texttt{.conf}} (configuration files), {\small\texttt{.jar}} (Java Archive file), and {\small\texttt{.so}} (libraries) are ruled out from being considered because they may be used consistently over multiple executions with the same names even with complex names. Second, for the rest of the filenames, if they exhibit the characteristic of frequent transitions among letters, digits, and special characters, they will be recognized as a hash-like identifier.
%Due to space limitations in the main text, we have included more detailed descriptions of this algorithm in the Appendix.

%for the rest of the files, we use frequent transitions of letters among alphabets, numeric characters, and special characters. 

In a nutshell, our normalization algorithms opportunistically generalize the names of non-representative file and directory names based on operating system knowledge and data with simple pattern matching. If their names are not matched, simply the raw names will be used without being replaced by generalized names. They would not cause runtime errors or incorrect results.

\subsection{Embedding Generation} 
The provenance record contains a large amount of unstructured data. To facilitate subsequent machine learning tasks, embedding methods are required. In natural language processing, embedding refers to the process of representing text as a fixed-length numerical vector in a continuous vector space. We have observed that multiple embedding techniques have been proposed for words~\cite{mikolov2013distributed}, sentences~\cite{pagliardini2018} and documents~\cite{le2014distributed}. 
This vector representation not only extracts the semantic meaning of the text but also can be further fed into a neural network. From the semiotics extraction stage, we have obtained textual descriptions for each system event. 
Therefore, in the embedding generation stage, we particularly adopt \texttt{Sent2Vec}~\cite{pagliardini2018} to generate numeric representations for these events.
%because in previous processing, we abstracted system events into sentences.
This method leverages $n$-gram features and an efficient unsupervised objective to capture the contextual information of sentences. As a result, we use a 50-dimensional vector to represent an event. Higher dimensions often have the capacity to encode richer information, but they also typically entail greater computational complexity. Considering the relatively small vocabulary size and the short sentence length in our application, we empirically choose 50 as the dimension of embeddings.

\input{pics/tab_dataset}

\subsection{Similarity Comparison} 
In Few-Shot Learning, Siamese networks stand out particularly in the area of detecting similarities among multiple comparable items, and they have already been applied in security applications\cite{fu2024seeing,zuo2019exploiting}. We thus propose a Siamese-network-based threat detection method that converts the anomaly detection problem into a semantic comparison problem. In the training phase, we prepare three types of event pairs: \ding{182} both events in a pair are benign; \ding{183} both events in a pair are adversary; \ding{184} one event is benign while the other is adversarial. Benign and adversary events inherently come from two different clusters in the feature space, and the intra-cluster distance is relatively larger than the inter-cluster distance. Consequently, the ground truth assigned to the first and second types of event pairs is a binary label false, i.e., the two events in a pair are considered similar. By contrast, the ground truth assigned to the third type of event pair is true, indicating that the two events in the pair are from two distinct clusters. 

Furthermore, we adopt a lightweight and shallow network design for each sub-network of the Siamese architecture, comprising three layers of fully-connected feed-forward neural networks with ReLU activation functions.
\begin{comment}
\begin{align*}
\rightarrow Flatten &\rightarrow linear(50, 128) \rightarrow ReLU \\
&\rightarrow linear(\_\_, 128) \rightarrow ReLU \\
&\rightarrow linear(\_\_, 128) \rightarrow ReLU 
\end{align*}
\end{comment}
%where 128 is the output size of each layer. 
The purpose herein is to explore how well the threat detector performs even when it only uses a simple network design. 
%When training the neural network, we also use a dropout rate of 0.1 to prevent overfitting. 
Once the neural network with fine-tuned weights is constructed through training, we can determine if a previously unseen event is adversarial or not by evaluating the discrepancy between it and a known event.

The optimization objective of our model training is to minimize a contrastive loss~\cite{hadsell2006dimensionality}, which is a distance-based loss. This loss function is widely used in few-shot learning to describe the degree of similarity between two samples. Namely, two similar inputs have a small distance and two dissimilar inputs instead have a relatively large distance. 
\begin{comment}

The contrastive loss is described as formula (\ref{eq_loss})
\begin{equation}\label{eq_loss}
(1-Y)\frac{1}{2}(\mathcal{D}_W)^2 + (Y)\frac{1}{2}\{max(0,m-\mathcal{D}_W)\}^2
\end{equation}
where $\mathcal{D}_W$ represents the Euclidean distance between the outputs of the two sub-networks in a Siamese architecture, and $Y$ is a binary label assigned to the input pair. Lastly, $m > 0$ is a margin used to define a radius around the output of one of the sub-networks. 
    
\end{comment}

Our evaluation shows that, even with a relatively small training dataset and a network with very few layers, our threat detection solution can still distinguish previously unseen malicious incidents from benign ones.

%% file: pics/alg1.tex
\begin{algorithm}[t]
\caption{An algorithm for normalization of a token}
\label{alg:normalization}
\begin{algorithmic}[1]
\Require token $t$, and {\small\texttt{IsHighVarietyCharacterName}}($t$)
\If{$t$ is in the directories reserved for temporary files or has the \texttt{.tmp} extension}
    \State return {\small\texttt{<TMP>}}
\EndIf
\If{$t$ is in the OS internal state directories such as {\small\texttt{/run/}}, {\small\texttt{/dev/}}, or {\small\texttt{/proc/}} directory}
    \State return {\small\texttt{<TMP>}}
\EndIf
\If{$t$ is a pipe}
    \State return {\small\texttt{<PIPE>}}
\EndIf
\If{$t$ is a hash-like identifier} 
    \State return \texttt{<HASH>}
\EndIf
\State return $t$
\end{algorithmic}
\end{algorithm}

%is a function to recognize names with a high variety of characters to be explained.
%{\small\texttt{IsHighVarietyCharacterName}}($t$) == {\small\texttt{True}}

%% file: pics/tab_dataset.tex
\begin{table*}[ht]
  \caption{The number of system events extracted from provenance data}\label{tab:db_sta}
   \centering
  \begin{center}
  
\begin{tabular}{|l|c|c|c|c|c|c|c|}
\hline
                                                         & \texttt{C01}$\sim$\texttt{C05} & \texttt{C06} & \texttt{C07} & \texttt{C08} & \texttt{C09} & \texttt{C10} & \texttt{C11}   \\ \hline
Events in \textit{ben.} scenarios                               & 4,341   & 272 & 324 & 1,620 & 336 & 339 & 3,678 \\ \hline
Events in \textit{adv.} scenarios                            & 4,516   & 438 & 237 & 2,527 & 477 & 399 & 5,041 \\ \hline
Overlap between \textit{ben.} and \textit{adv.}   & 2,629   & 178 & 190 & 1,202 & 309 & 324 & 2,958 \\ \hline
Events in \textit{adv.} after refinement & 1,887   & 260 & 47  & 1,325 & 168 & 75  & 2,083 \\ \hline

\end{tabular}
\end{center}
\end{table*}

%% file: eval.tex
\section{Evaluation}\label{sec:eval}

%In this section, we illustrate the experiment settings, 
%visualize the initial event embeddings through manifold learning, 
%and 
%empirically 
%evaluate the effectiveness of the proposed technique. 

%\input{pics/tab_vul}

\subsection{Dataset}

The performance of our technique was evaluated using a publicly accessible research dataset~\textsc{ProvSec}\cite{Shrestha2023provsec}, which includes provenance analysis data collected from a cloud environment involving 11 attack scenarios. 
They reflect real cyber attacks in cloud based on CVE entries and the corresponding proof-of-concept (PoC) exploit code. In the following description, we use \texttt{C01}$\sim$\texttt{C11} to represent each category of attacks.
%Intuitively, the system events from the entire target machine are recorded as  ``benign'', if the corresponding workload does not involve any attacks. In contrast, the system events for the workload under an attack are marked as ``adversary''. 
%Table~\ref{tab:vuls} illustrates the 11 attack scenarios that are evaluated in our experiments. 
%These cases reflect real cyber attacks in cloud based on CVE entries and the corresponding proof-of-concept (PoC) exploit code. 

When the attack is on, the corresponding provenance data of a specific adversary scenario is recorded. We accordingly extract events and preliminarily consider them as adversarial. Otherwise, if the attack has not been launched, the events extracted from the corresponding provenance data are regarded as benign. It is unsurprising that there exist some events that may be on both sides. The intersection between the benign events and the preliminary adversary events is not sufficiently iconic to indicate an attack behavior. Therefore, we remove those overlapped patterns from the original adversary events to obtain a refined dataset. The details of events per category extracted from provenance data are shown in Table~\ref{tab:db_sta}.

Furthermore, we prepare both similar and dissimilar pairs in the dataset to train a Siamese-network-based model. The two events in a dissimilar pair are from different clusters, i.e., one is benign while the other one is adversarial. In contrast, the two events in a similar pair are from the same clusters. Namely, they must be either both benign or both adversarial. Table~\ref{tab:db_train} shows the number of pairs per category, where the sizes of similar and dissimilar pairs are balanced. Additionally, among the similar pairs, half are benign pairs while the other half are adversarial.

\subsection{Evaluation Metrics}\label{metrics}

\input{pics/tab_pairs}

$Precision$, $Recall$, $F_1\ score$, and $Accuracy$ are commonly used metrics in the field of machine learning to evaluate the performance of classification models. $Precision$ is a measure of how many of the positively predicted instances are actually true positives. High precision indicates that the model has a low rate of false positives. False positive is the number of predicted similar pairs that actually are not. $Recall$, aka True Positive Rate (TPR), is the measurement describing how robust the model is in identifying malicious attacks. A high $Recall$ means the method can effectively distinguish a benign event from an adversary one. $F_1\ score$ is the harmonic mean between $Precision$ and $Recall$. A higher $F_1\ score$ implies a lower false positive rate as well as a lower false negative rate. $Accuracy$ is the ratio of the number of correct predictions to the total number of input samples. Thus, a high $Accuracy$ means that the model performs well overall.

\begin{figure*}
     \centering
     \begin{subfigure}[b]{0.3\textwidth}
         \centering
         \includegraphics[width=\textwidth]{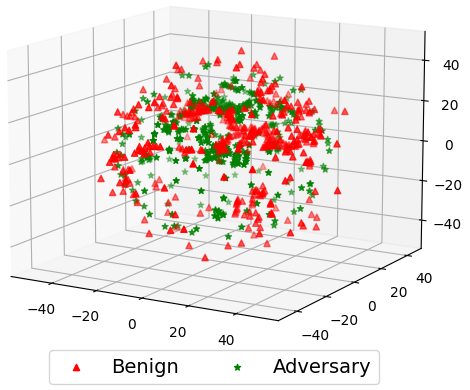}
         \caption{\texttt{D01}$\sim$\texttt{D05}}
         \label{fig:001}
     \end{subfigure}
     \hfill
     \begin{subfigure}[b]{0.3\textwidth}
         \centering
         \includegraphics[width=\textwidth]{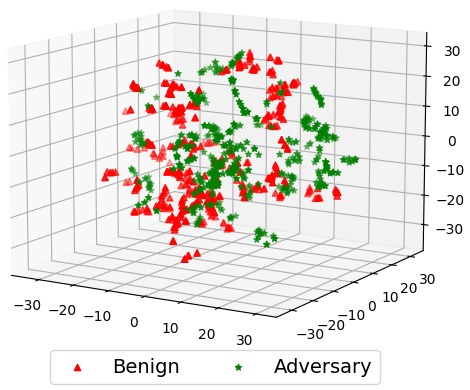}
         \caption{\texttt{D08}}
         \label{fig:002}
     \end{subfigure}
     \hfill
     \begin{subfigure}[b]{0.3\textwidth}
         \centering
         \includegraphics[width=\textwidth]{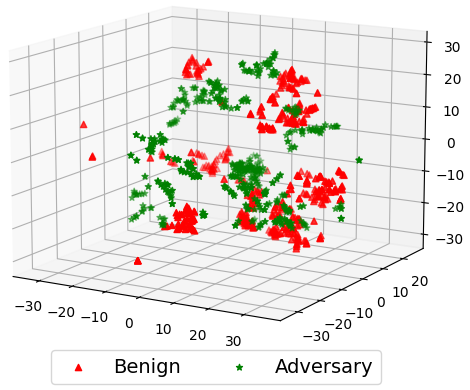}
         \caption{\texttt{D11}}
         \label{fig:003}
     \end{subfigure}
        \caption{Visualization of initial event embeddings based on \texttt{Sent2Vec}~\cite{pagliardini2018}. }
        \label{fig:three graphs}
\end{figure*}
%The coordinate axes respectively represent three dimensions obtained by \texttt{t-SNE}~\cite{van2008visualizing}.
\subsection{Embedding Visualization and Model Interpretability}

Intuitively, if the numerical representations for events provided by the embedding model are semantically meaningful, the degree of similarity between events can be often reflected in the embedding space by the proximity or distance. Therefore, we examine the interpretability of event embeddings generated by \texttt{Sent2Vec}~\cite{pagliardini2018} through visualization. In detail, we randomly selected 1,000 samples from \texttt{D01}$\sim$\texttt{D05}, \texttt{D08}, and \texttt{D11}, respectively. The reason we choose these three categories is that there are too few samples in the other sets. We project the event embeddings to a three-dimensional space using \texttt{t-SNE}~\cite{van2008visualizing}. As Figure~\ref{fig:three graphs} shows, the red points represent benign samples while and green ones represent adversarial samples. It should be noted that the numbers of benign and adversarial samples are balanced. It turns out that benign and adversarial samples are not trivially separable, especially when mixing different attacks together, as Figure~\ref{fig:three graphs}(a) shows. One explanation is that we compress the high-dimensional space into a three-dimensional space, and due to information loss, the separating hyper-planes between groups that could originally be distinguished are no longer apparent. On the other hand, we can still observe small cliques formed by similar events, which is particularly evident in Figure~\ref{fig:three graphs}(c). This implies that the existing event embeddings can be used as a feasible starting point. Based on this, we can take advantage of the subsequent FSL-based neural network model to further amplify the distance between benign and adversary events in a feature space. At the same time, the distance between two benign events, and the distance between two adversarial events can both be reduced.

\subsection{Effectiveness}

As previously mentioned, to address real-world cyber incidents in cloud environments, we are particularly interested in an effective threat detection system capable of detecting previously unseen APT attacks. 
%An important intuition we have is different APT attacks follow a similar pattern, as confirmed by existing research (See Section 3). %Therefore 
Ideally, the knowledge acquired from a few known attack scenarios is supposed to generalize over other unseen attacks. 
To validate our hypothesis, we divided the data into 7 subsets based on different attacks, as shown in Table~\ref{tab:db_train}. 
After that, we conducted four groups of experiments, from Group 1 to Group 4, where the size of the training set gradually increases in each group. Models trained on known attacks will be used to test unseen attacks.
%, as shown in Table~\ref{tab:eval}. 
Specifically, we will use the metrics described in Section~\ref{metrics} to evaluate the effectiveness of the attack detection models.

We perform the evaluation in multiple groups as we change the size of the training set and the testing set. Table~\ref{tab:eval} summarizes our main results. Group 1 uses five cases from \texttt{D01} to \texttt{D05}. We also varied the group by including additional data cases in the training set. Groups 2, 3, and 4 respectively have six, seven, and eight cases in the training set. The remaining unused cases in the training stage are evaluated as the testing set. The FSL-based method we proposed has shown promising results overall, achieving an average accuracy of 91.7\% across 18 tests.

%We believe this is a new attempt to apply FSL to system call behavior extracted from provenance data. Although several existing methods for threat detection using provenance analysis have involved system call behavior, previous approaches either utilized provenance graphs or extracted paths from these graphs for detection or analysis. In contrast, our focus is on a granularity at the level of system events. We could not identify any suitable and solid work for comparison, whose methods are also based on a granularity similar to that of system events. Thus, we present our evaluation as we vary the sizes of training sets, so that the smallest training set can be a baseline case to be compared with larger training set cases.

\subsection{Case Study}

In general, we observed the FSL model performs differently depending on the evaluated data cases as it has more cases in the training set. Some cases like \texttt{D09} have better accuracy, precision, and recall in the Group 4 experiment with 8 cases in the training. This is typically in line with our expectations, \textit{``as the model sees more data, it can predict better''}. 
%However, we found this is not always the case. \texttt{D10} performs better in Group 1. Its performance does not show a clear pattern with the growth of the training set. 

\input{pics/tab_eval}

%(Insight) 
Also, we noticed that the performance on the \texttt{D10} category is often worse than that on other categories. 
Our manual examination found that the \texttt{D10} case, an SQL injection case, exhibits more distinct behavior compared to the other attack cases causing less consistent behavior.
%due to XXX.
%(although it is not always as seen in D10 of Group 4).
%Figure \ref{fig:d10} shows the provenance graph of the \texttt{D10} case. The python process whose process ID is 43739 shows its behavior to access the \texttt{hosts} file and \texttt{technical\_500.html} file for HTTP error handling which are very specific operations unique to this exploit. 
Other cases include similar behavior, such as command injections and new shell invocations due to the shell code, which are represented as new program invocations even though they are performed by different programs with different exploits. This unique behavior of this particular case made its behavior far from other cases causing unusual behavior.

\begin{comment}

\begin{figure}[!ht]
  \centering
  \includegraphics[width=0.8\textwidth]{pics/10-a.scap.tscap_bt.pdf}
  \caption{The provenance graph of the \texttt{D10} 
 exploitation of Django's SQL injection vulnerability (CVE-2021-35042). This case does not involve the patterns commonly seen in command injections or new shell invocations.}
  \label{fig:d10}
\end{figure}

%(Clarification on missing evaluation compared with related work as baselines) 

\end{comment}

%% file: pics/tab_pairs.tex
\begin{table}[!th]
  \caption{The datasets consisting of similar and dissimilar event pairs}\label{tab:db_train}
     \centering
  \begin{center}
\begin{tabular}{@{}c|ccccccc@{}}
\hline
Dataset & \texttt{D01}$\sim$\texttt{D05} & \texttt{D06} & \texttt{D07} & \texttt{D08}  & \texttt{D09} & \texttt{D10} & \texttt{D11}  \\ \hline

Source  & \texttt{C01}$\sim$\texttt{C05} & \texttt{C06} & \texttt{C07} & \texttt{C08}  & \texttt{C09} & \texttt{C10} & \texttt{C11}  \\

\# of pairs    & 3,380    & 516 & 92  & 2,648 & 332 & 148 & 4,164 \\ \hline
\end{tabular}
\end{center}
\end{table}

%% file: pics/tab_eval.tex
% \usepackage{multirow}
\begin{table}[h!]

\caption{Evaluation on unseen attacks}\label{tab:eval}
     \centering
  \begin{center}
\begin{tabular}{|c|c|c|c|c|c|c|}
\hline
                      Group & Train & Test & Accuracy & Precise & Recall & F$_1$  \\ \hline
\multirow{6}{*}{G1} & \texttt{D01$\sim$D05}      & \texttt{D06}         & 88.2\%   & 93.0\%  & 82.6\% &  87.5\%  \\ \cline{2-7} 
                      & \texttt{D01$\sim$D05}      & \texttt{D07}         & 91.3\%   & 95.2\%  & 87.0\% &  90.9\%  \\ \cline{2-7} 
                      & \texttt{D01$\sim$D05}      & \texttt{D08}         & 92.9\%   & 88.9\%  & 98.2\% &  93.3\%  \\ \cline{2-7} 
                      & \texttt{D01$\sim$D05}      & \texttt{D09}         & 94.0\%   & 90.1\%  & 98.8\% &  94.3\%  \\ \cline{2-7} 
                      & \texttt{D01$\sim$D05}      & \texttt{D10}         & 85.8\%   & 87.3\%  & 83.8\% &  85.5\%  \\ \cline{2-7} 
                      & \texttt{D01$\sim$D05}      & \texttt{D11}         & 94.7\%   & 91.2\%  & 99.0\% &  95.0\%  \\ \hline
\multirow{5}{*}{G2} & \texttt{D01$\sim$D06}      & \texttt{D07}         & 91.3\%   & 93.2\%  & 89.1\% &  91.1\%  \\ \cline{2-7} 
                      & \texttt{D01$\sim$D06}      & \texttt{D08}         & 95.1\%   & 92.1\%  & 98.7\% &  95.3\%  \\ \cline{2-7} 
                      & \texttt{D01$\sim$D06}      & \texttt{D09}         & 95.5\%   & 98.8\%  & 92.7\% &  95.6\%  \\ \cline{2-7} 
                      & \texttt{D01$\sim$D06}      & \texttt{D10}         & 82.4\%   & 87.5\%  & 75.7\% &  81.2\%  \\ \cline{2-7} 
                      & \texttt{D01$\sim$D06}      & \texttt{D11}         & 96.2\%   & 94.3\%  & 98.3\% &  96.3\%  \\ \hline
\multirow{4}{*}{G3} & \texttt{D01$\sim$D07}      & \texttt{D08}         & 93.2\%   & 88.7\%  & 98.9\% &  93.5\%  \\ \cline{2-7} 
                      & \texttt{D01$\sim$D07}      & \texttt{D09}         & 94.0\%   & 89.2\%  & 100\%  &  94.3\%  \\ \cline{2-7} 
                      & \texttt{D01$\sim$D07}      & \texttt{D10}         & 83.8\%   & 85.7\%  & 81.1\% &  83.3\%  \\ \cline{2-7} 
                      & \texttt{D01$\sim$D07}      & \texttt{D11}         & 96.0\%   & 93.4\%  & 98.9\% &  96.1\%  \\ \hline
\multirow{3}{*}{G4} & \texttt{D01$\sim$D08}      & \texttt{D09}         & 96.4\%   & 93.3\%  & 100\%  &  96.5\%  \\ \cline{2-7} 
                      & \texttt{D01$\sim$D08}      & \texttt{D10}         & 83.1\%   & 90.2\%  & 74.3\% &  81.5\%  \\ \cline{2-7} 
                      & \texttt{D01$\sim$D08}     & \texttt{D11}         & 96.6\%   & 95.3\%  & 98.0\% &  96.6\%  \\ \hline
\end{tabular}
\end{center}
\end{table}

%% file: relate.tex
\section{Related Work}\label{sec:related}

%In this section, we briefly present a literature review. Considering the large volume of research in the related fields, this review is not intended to be exhaustive.

\subsection{Provenance Analysis}
Using provenance in intrusion analysis and detection has been explored by a large body of work
\cite{zipperle2023survey}.
Dependence tracking analysis \cite{king2003backtracking} 
has been used to analyze a large volume of data effectively.
Provenance tracking has been done in different data granularity.
BEEP \cite{beep} and 
\textsc{ProTracer} \cite{protracer} use units that are execution partitions of application code which is common in event-handling loops.
MPI \cite{mpi} uses  user input on data structures to define execution partitions.
\textsc{PrioTracker} \cite{priotracker} proposed priority-based causality tracking using rareness score and fanout score as indications of unusualness.
Bates et al. \cite{whole} proposed Linux Provenance Modules (LPM), a kernel-based framework and data loss prevention system for sensitive data.
\textsc{PalanTir} \cite{palantir} uses a processor tracing (PT) hardware technique to enable finer-grained instruction level tracking.
% recent work
\textsc{Kairos} \cite{cheng2024kairos} proposed a graph neural network-based encoder and decoder to learn the temporal evolution of the provenance graph's structural changes.
We analyze the provenance data using few-shot learning algorithm with the goal of detecting anomalies from the behavior of few samples.

\subsection{Attack detection}
Using provenance data for attack detection is a branch of active research.
Multiple attack detection approaches have been proposed based on dependency graphs.
\textsc{Holmes} \cite{homes} proposed an approach to detect and summarize APT attack campaign stages.
\textsc{Sleuth} \cite{203676} proposed a tag-based attack detection system to prioritize behavior. 
Hossain et al. \cite{9152772} improved a tag-based system reducing false alarms significantly in the detection of APT-style attacks.
\textsc{NoDoze} \cite{nodoze} uses a network diffusion algorithm that propagates anomaly scores across dependency graphs to calculate anomaly scores. Hassan et al. \cite{9152771} propose another graph-based scoring scheme for an alert triage system with path preferences and graph reduction schemes. In this work, we propose using Few-Shot Learning on the provenance data due to the growing diversity of attack mechanisms. The presented results shed light on a promising direction for scenarios where adversary data is inadequate.

%% file: conclude.tex
\section{Conclusion}\label{sec:conclude}

System provenance analysis is believed to be a promising mechanism. Due to its remarkable ability to track  dependencies across system events in a cyber incident, this potent technique can play a crucial role in the analysis of incidents as well as in detecting attacks. While various attack patterns are possible, having a limited set of attack datasets becomes a challenge.
In this paper, we propose a new approach that applies few-shot learning to attack detection based on system events. Our evaluation of a set of 11 different attack scenarios shows a promising result of 91.7\% accuracy on average from 18 experiments with varied training and testing sets, demonstrating that the threat prediction on different attacks with few data samples is possible.

%% file: ack.tex
\section*{Acknowledgment}

Sandia National Laboratories is a multimission laboratory managed and operated by National Technology and Engineering Solutions of Sandia, LLC., a wholly owned subsidiary of Honeywell International, Inc., for the U.S. Department of Energy’s National Nuclear Security Administration under contract DE-NA0003525. This article describes objective technical results and analysis. Any subjective views or opinions that might be expressed in the article do not necessarily represent the views of the U.S. Department of Energy or the United States Government. This work was supported through contract CR-100043-23-51577 with the U.S. Department of Energy.

\begin{comment}

\section*{Acknowledgment}

The preferred spelling of the word ``acknowledgment'' in America is without 
an ``e'' after the ``g''. Avoid the stilted expression ``one of us (R. B. 
G.) thanks $\ldots$''. Instead, try ``R. B. G. thanks$\ldots$''. Put sponsor 
acknowledgments in the unnumbered footnote on the first page.

\end{comment}